# Formation of hexagonal Boron Nitride on Graphene-covered Copper Surfaces

Devashish P. Gopalan,[1] Patrick C. Mende,[1] Sergio C. de la Barrera,[1] Shonali Dhingra,[2] Jun Li,[1] Kehao Zhang,[3] Nicholas A. Simonson,[3] Joshua A. Robinson,[3] Ning Lu,[4] Qingxiao Wang,[4] Moon J. Kim,[4] Brian D'Urso,[2] Randall M. Feenstra[1,a]

[1]Department of Physics, Carnegie Mellon University, Pittsburgh, PA 15213
[2]Department of Physics and Astronomy, University of Pittsburgh, Pittsburgh, PA 15260
[3]Department of Materials Science and Engineering and Center for Two-Dimensional and Layered Materials, The Pennsylvania State University, University Park, PA 16802
[4]Department of Materials Science and Engineering, The University of Texas at Dallas, Richardson, TX 75080

## Abstract

Graphene-covered copper surfaces have been exposed to borazine, $(BH)_3(NH)_3$, with the resulting surfaces characterized by low-energy electron microscopy. Although the intent of the experiment was to form hexagonal boron nitride (h-BN) on top of the graphene, such layers were not obtained. Rather, in isolated surface areas, h-BN is found to form μm-size islands that substitute for the graphene. Additionally, over nearly the entire surface, the properties of the layer that was originally graphene is observed to change in a manner that is consistent with the formation of a mixed h-BN/graphene alloy, i.e. h-BNC alloy. Furthermore, following the deposition of the borazine, a small fraction of the surface is found to consist of bare copper, indicating etching of the overlying graphene. The inability to form h-BN layers on top of graphene is discussed in terms of the catalytic behavior of the underlying copper surface and the decomposition of the borazine on top of the graphene.

## I. Introduction

A large number of growth studies have been conducted over the past decade for various two-dimensional (2D) materials including graphene and hexagonal boron nitride (h-BN), with metal substrates being often employed.[1,2,3] The presence of the metal is generally acknowledged to provide some catalytic activity for the decomposition of the precursor molecules and the subsequent formation of the graphene or h-BN films.[4,5] Indeed, for single monolayer of h-BN, this growth mode was elucidated in the early works of Nagashima et al. and Auwärter et al.[6,7,8]

For application in electronic devices, thin films of 2D materials must be removed from those substrates and then transferred onto an insulating material.[9,10] For *heterostructures*, containing thin layers of different materials, the number of steps needed to build up the structure can be relatively large. Possible contamination induced by the transfer process, for each transfer step, might then be deleterious to the electrical properties of the final device.[11,12] For this reason, a number of authors have investigated the growth of different 2D layers, one on top of the other, with the goal of epitaxially forming a heterostructure.[13,14]

For the case of graphene and h-BN, prior studies have been performed for both graphene on h-BN and h-BN on graphene.[13,15] The former is somewhat more common, since h-BN itself has been demonstrated to be a relatively ideal, insulating substrate for growth (or transfer) of 2D layers.[16] Nevertheless, some work involving h-BN on graphene has been reported.[17] In principle, the ability to deposit h-BN layers on graphene, and then follow that by a subsequent graphene deposition, could lead to the formation of large-area graphene-insulator-graphene (GIG) tunneling junctions. Such devices have recently been demonstrated to produce highly nonlinear current-voltage characteristics, with application for high-speed transistors, oscillators, and other novel devices.[18,19,20,21]

---

[a] feenstra@cmu.edu

We have recently completed a study of the growth of h-BN layers on *epitaxial graphene* formed on SiC surfaces, using a borazaine precursor.[22] In that work, for sample temperatures near 1100°C, we obtained ~2 μm-size islands of the h-BN on the graphene, with preferential epitaxial orientation between the h-BN and the graphene. Building on this prior work, we have in the present study attempted the growth of h-BN on graphene, but now using graphene on Cu as a starting substrate. We find that the results are quite different than for the h-BN growth on epitaxial graphene on SiC, in that we do not observe any h-BN layers formed on top of graphene. Rather, we observe isolated surface areas in which h-BN is found to form μm-size islands that *substitute* for the graphene. Additionally, over the majority of the surface, the properties of the layer that was originally graphene is observed to change what appears to be an h-BNC alloy.[3] Furthermore, a small fraction of the surface is found to etched by the borazine, resulting in bare, exposed copper.

The main characterization tool we employ to study our surfaces is the low-energy electron microscope (LEEM).[23] This instrument allows real-space imaging at low electron energies (typically 0 – 20 eV) as well as low-energy electron diffraction (LEED) capability at selected, μm-size surface locations (μLEED). Additionally, by acquiring a sequence of images as a function of energy, one can extract low-energy electron reflectivity (LEER) spectra. Such spectra provide a unique "fingerprint" of the local electronic structure of the surface being probed,[24] thereby yielding chemical information about the surface composition. In this work we also develop a quantitative method whereby relative work functions can be extracted from the data, providing an additional means of characterizing the electronic structure of the surface.

Over the past several years we have conducted a wide range of LEEM studies of 2D materials on various substrates. From the assortment of LEER spectra thus acquired, we are able to interpret new spectra from surfaces having unknown structures and hence learn about the structure of the surfaces in question. Additionally, over the same time period we have developed a simulation capability for LEER spectra.[25,26,27] Given some specific surface structure, we perform a computation of its LEER spectrum, and then by comparing experiment and theory we can deduce whether or not the assumed surface structure matches the one in experiment.

Figure 1 shows a collection of LEER spectra from surfaces that we have recently studied. All of these surfaces are described in detail in separate publications;[22,28,29] we display them here to illustrate our method of surface identification based on LEER. First, considering Figs. 1(a) – 1(c), these are from our above-mentioned study of h-BN on epitaxial graphene. For the first two spectra, acquired from 1 and 2 monolayers (MLs) of graphene, they reveal one or two distinct minima, respectively, in the low energy range of 0 – 5 eV. Such spectra for epitaxial graphene are well understood based on recent studies.[25,26,30] The reflectivity minima arise from *interlayer states*, which are plane-wave type states that form in the spaces *between* graphene layers. In general $n$ layers of a 2D material will have $n-1$ spaces between the layers. Hence, $n-1$ combinations of interlayer states are formed and these lead to $n-1$ minima in a LEER spectrum. For the specific case of epitaxial graphene on SiC, there is an additional, underlying graphene-like layer on the surface, the so-called buffer layer,[31] and an interlayer state is also formed between that layer and the graphene ones above. Hence each minimum in the LEER spectra corresponds to one graphene layer (as first deduced by Hibino et al.[30]).

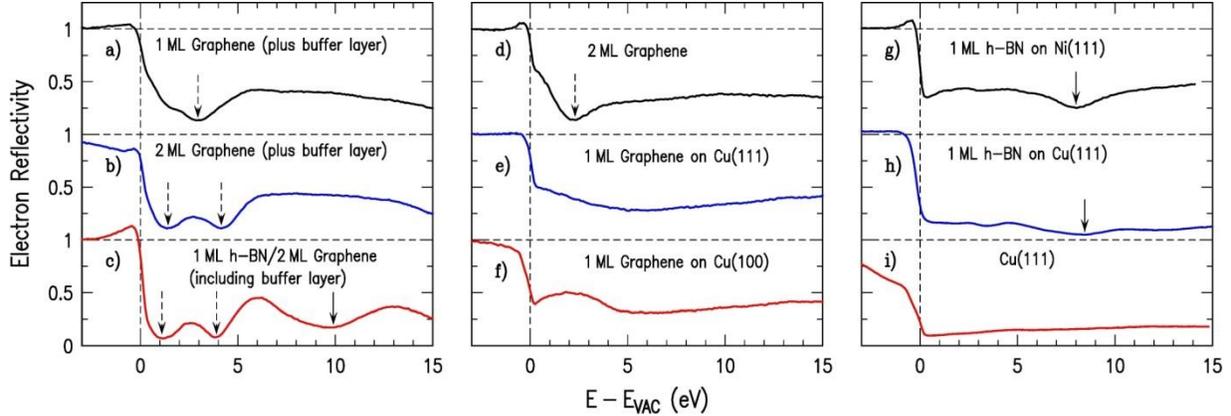

FIG 1. LEER spectra obtained from a variety of surfaces, as labelled: (a) – (c): epitaxial graphene on SiC, with 0 or 1 h-BN layers on top of the graphene; (d) – (f): CVD-grown graphene on Cu; (g) – (i): h-BN on Ni and Cu substrates. Downwards pointing *dotted* arrows indicate interlayer states, and *solid* arrows indicate portions of the spectra that derive from additional h-BN electronic states.

When h-BN is present on the surface, then the situation changes. The interlayer states and associated reflectivity minima still exist in the 0 – 5 eV range, as just described, but the h-BN produces an additional minimum located at 8 – 9 eV. This additional minimum is formed from a specific band structure feature of the h-BN that, again, is well understood based on recent work.[22,27] This feature can be seen in Fig. 1(c), as well as Figs. 1(g) and 1(h) for h-BN on Ni and Cu, respectively, and it permits identification of h-BN on the surface. (In Figs. 1(g) and 1(h), a very small oscillation in the reflectivity is also apparent at about 4 eV, and, like the minimum at 8 – 9 eV, this small feature is also characteristic of the h-BN).[22]

Now considering the situation when graphene or h-BN resides directly on a metal surface, then as previously discussed,[26] an interlayer state can form between the 2D layer and the surface so long as the separation between the two is sufficiently large, ≥3 Å. The energy of this interlayer state varies inversely with the separation. For single-layer graphene on Cu(111) and Cu(100) surfaces, as seen in Figs. 1(e) and 1(f), the broad minima centered at about 6 eV arise from this interlayer state. For Fig. 1(d), with 2 ML of graphene, the dominant feature is simply the distinct interlayer state arising from the space between the two graphene layers, and the underlying interlayer state (from the graphene-Cu space) is scarcely visible. Finally, for Fig. 1(f) an additional feature is visible, a plateau extending from about 0 – 4 eV. This feature arises from the band structure of the underlying Cu(100) substrate, which has a bandgap in its energy spectrum for electrons propagating in the (100) direction.

We will return to discussion of LEER spectra when we present additional results in Section III. The remainder of this paper is organized as follows. Experimental details are provided in Section II, including our method for extracting work function variations over the surface. Section III details our observations of the effects of borazine exposure for both relatively low sample temperature (900°C) and high temperature (1000°C). A discussion of the results is given in Section IV. The apparent inability to form h-BN layers on top of graphene is discussed there in terms of the catalytic behavior of the underlying copper surface and the decomposition of the borazine on top of the graphene.

## II. Experimental Methods

Graphene growth was achieved using atmospheric pressure chemical vapor deposition (APCVD) on ultra-flat copper substrates. To prepare these substrates, bulk oxygen-free electronic grade ultra-pure (99.99%) copper rods were used as starting material. These 30 cm copper rods, 25.4 mm in diameter, were then machined down to ~1.2 mm thick slices, using conventional machining tools and single point diamond turning. Before graphene growth, these substrates were annealed for 8 hours at 1000°C, in 70 sccm of 2.5 vol % $H_2$/Ar mixture. During the growth process, the substrates were subsequently annealed at 1050°C for 1 hour under 186 sccm flow of 2.5 vol % $H_2$/Ar mixture. 14 sccm of 0.1 vol % $CH_4$/Ar mixture was then introduced for 1.5 hours as the precursor gas. This procedure has shown to produce large continuous 1 ML thick graphene domains.[32,33] The ultra-flat copper substrates used for this process were shown to have a root mean square surface roughness of ~2 nm, resulting in graphene that is ~50 times smoother than graphene obtained on standard 25 μm thick copper foils.[32] The samples were then characterized using LEED and LEEM prior to h-BN growth.

Hexagonal BN growth was carried out in a high-vacuum deposition system, with base pressure of $1\times10^{-9}$ Torr. After degassing, the samples were exposed to $10^{-4}$ Torr of borazine, $(BH)_3(NH)_3$, for 30 minutes. During borazine exposure, the sample was heated to temperatures of 900°C or 1000°C. As will be discussed later, the surface morphology is governed by the growth temperature. In brief, the samples prepared at 900°C mainly retain the 1-ML-graphene coverage of the starting substrate, as well as forming some h-BN islands, whereas for the samples prepared at 1000°C the original graphene is nearly all converted to h-BNC alloy.

Immediately following borazine exposure, the samples were transferred in situ to an Omicron SPECTALEED (large area) low-energy electron diffraction (LEED) system, which also allowed in-situ measurement of Auger electron spectroscopy (AES). Further characterization was performed by removing the samples from the growth system and transferring them through air to an Elmitech III low-energy electron microscope (LEEM). This system also contains a VG Scientific Clam 100 hemispherical analyzer which was employed for ex-situ AES measurements (5 keV electrons). Additional characterization was performed with a JEOL 2100F transmission electron microscope (TEM) with electron beam energy of 200 keV, utilizing a GIF Tridiem 863 system for electron energy loss spectroscopy (EELS) and mapping. Prior to this measurement, the BN/graphene films were separated from the Cu substrate and transferred onto a TEM grid.

In the LEEM, the samples were outgassed by heating to ~300°C for 20 minutes. In addition to providing information on the structure and stoichiometry of the surface, the LEER spectra can be used to determine the difference in work function, $\Delta W \equiv W_S - W_C$, between the sample surface and the electron emitter, or cathode, of the LEEM (the emitter is $LaB_6$, which has a relatively low work function). We describe here a method that we have developed for quantitatively obtaining the $\Delta W$ values from the data.

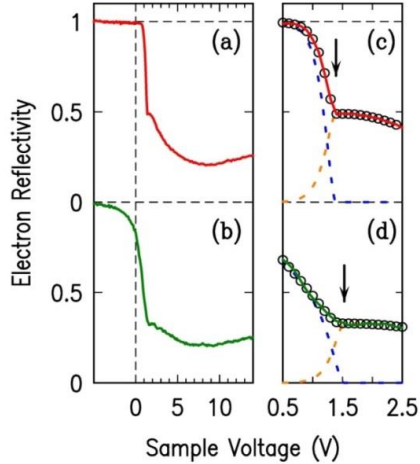

FIG 2. (a) and (b) Typical LEER spectra, with (a) displaying a sharp transition to unit reflectivity (near 1.5 V sample voltage) and (b) showing a more gradual transition. (c) and (d) Expanded views of the transition regions from panels (a) and (b), respectively. Black circles show a fit function, with the arrows indicating the onset voltages derived from the fit. The two components of the fits, for each spectrum, are indicated by the dotted lines.

Figures 2(a) and 2(b) show spectra obtained from different locations of our graphene on copper surfaces (the locations from which these spectra were acquired will be discussed in more detail in Section III(A)). The measurement is performed as a function of the sample voltage, $V$, which is the potential difference between the sample and the emitter,

$$eV = E_F^c - E_F^s \tag{1a}$$
$$= (E_{VAC}^s - E_F^s) - (E_{VAC}^c - E_F^c) + E_{VAC}^c - E_{VAC}^s \tag{1b}$$
$$= \Delta W + E_{VAC}^c - E_{VAC}^s \tag{1c}$$

where the Fermi energies of sample and emitter are denoted by $E_F^s$ and $E_F^c$, respectively, their vacuum levels by $E_{VAC}^s$ and $E_{VAC}^c$, and their work functions by $W_s \equiv E_{VAC}^s - E_F^s$ and $W_c \equiv E_{VAC}^c - E_F^c$.

For a relatively ideal spectrum such as in Fig. 2(a), we see, as a function of decreasing voltage, a sharp onset (near 1.5 V) at which the reflectivity rises to unity. This signifies the transition to "mirror mode" of the LEEM;[23,34] as pictured in Fig. 3(a), for sample voltage lower than this onset, the incident electrons do not have sufficient energy to reach the surface. Rather, they are reflected by the electric field (typically $10^4$ V/mm) that extends out from the surface to the objective lens of the electron optics. For a sample voltage equal to the onset voltage, the vacuum levels of the sample and emitter are aligned. Denoting the onset voltage by $V_0$, we have

$$eV_0 = \Delta W. \tag{2}$$

For voltages greater than the onset, all electrons are reflected from the sample surface or absorbed into the sample, as pictured in Fig. 3(b).

A convenient way to plot reflectivity spectra (as already employed in Fig. 1) is in terms of the energy of a sample state, as probed by the incident electrons. Electrons emitted from the thermionic emitter have well-known energy distribution, $N(\varepsilon) = \varepsilon \exp(-\varepsilon/\sigma_c)/\sigma_c^2$, with $\sigma_c = kT_c$ where $k$ is Boltzmann's constant and $T_c$ is the temperature of the emitter (cathode), and with $\varepsilon$ being the electron energy relative $E^C_{VAC}$.[35] This distribution is peaked at $\varepsilon = \sigma_c$, so that the incident electrons have peak energy of $\sigma_c + E^C_{VAC}$. This energy corresponds to the energy of a probed sample state, which we denote by $E$. Therefore, for plotting the spectra we employ

$$E - E^S_{VAC} = \sigma_c + E^C_{VAC} - E^S_{VAC} \tag{3a}$$
$$= \sigma_c + e(V - V_0) \tag{3b}$$

where the second line follows from the first by using Eqs. (1c) and (2). In our labelling of the spectral plots, we drop the superscript 's' from $E^S_{VAC}$, i.e. with $E - E_{VAC}$ understood to refer to the energy of a sample state relative to the vacuum level of the sample.

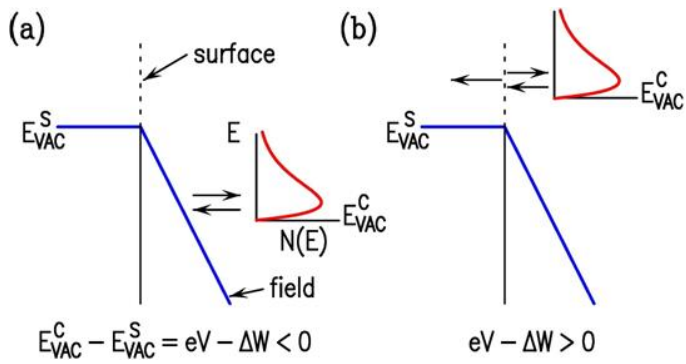

FIG 3. (a) and (b) Schematic energy diagrams of the distribution *N(E)* of electrons incident on the surface of a sample. In (a), the electrons are reflected by the field extending out from the surface, whereas in (b) the electrons have sufficient energy to reach the surface, where they are partially reflected and partially absorbed.

In order to obtain values for $\sigma_c$ and $V_0$ from the data, we employ a least-squares fitting procedure. Consider the situation of Fig. 3(a) with $V < V_0$; some electrons of the incident distribution will be reflected by the field. The number of those mirror-reflected electrons is given by $f_m(V) = \int_0^{\varepsilon_1} [\varepsilon \exp(-\varepsilon/\sigma_c)/\sigma_c^2] d\varepsilon$ where the upper limit of integration is $\varepsilon_1 = E^S_{VAC} - E^C_{VAC} = -e(V - V_0)$. Evaluating the integral, we find $f_m(V) = 1 - [1 - (V - V_0)/\sigma] \exp[(V - V_0)/\sigma]$ where $\sigma \equiv \sigma_c/e$. The number of electrons reflected from the sample is given by $f_s(V) = [1 - f_m(V)] r(E)$, where $r(E)$ is the reflectivity of the electrons at an energy given by Eq. (3b). Now considering the situation of Fig. 3(b) with $V > V_0$, we have no electrons being reflected by the field, $f_m(V) = 0$, and the number of electrons being reflected from the sample surface is given simply by $f_s(V) = r(E)$. For fitting the observed spectra, we do not assume that the data is necessarily normalized to unit reflectivity (i.e. at large, negative sample voltages). Hence, for the field-reflected electrons, we employ a fit function of the form

$$g_m(V) = \begin{cases} a_0 \{1 - [1 - (V - V_0)/\sigma] \exp[(V - V_0)/\sigma]\}, & V \leq V_0 \quad (4a) \\ 0, & V > V_0 \quad (4b) \end{cases}$$

where $a_0$ is a fit parameter. For the electrons reflected from the sample surface, we must assume some form for the reflectivity $r(E)$. We expand this function as a 2$^{nd}$ degree polynomial about an energy (relative to $E_{VAC}^s$) of $e(V-V_0)$, yielding the fit function

$$g_s(V) = \begin{cases} G_s \{[1-(V-V_0)/\sigma]\exp[(V-V_0)/\sigma]\}, & V \leq V_0 \quad (5a) \\ G_s, & V > V_0 \quad (5b) \end{cases}$$

where $G_s = b_0 + b_1(V-V_0) + b_2(V-V_0)^2$, with $b_0$, $b_1$, and $b_2$ all being fit parameters.

Thus, for a relatively ideal spectrum such as that of Fig. 2(a), we fit the data to $g_m(V) + g_s(V)$, with the fit employing the four linear parameters $a_0$, $b_0$, $b_1$, and $b_2$ along with the two nonlinear parameters $V_0$ and $\sigma$. The result is shown in Fig. 2(c), with best-fit values of $V_0 = 1.385 \pm 0.004$ V and $\sigma = 0.121 \pm 0.003$ V. We obtain a very good fit for a voltage window extending over $\pm 1$ V or more on either side of the onset, for spectra such as this, yielding a relative work function $\Delta W = eV_0$ with less than 10 meV uncertainty. The value obtained here for the width of the electron distribution, 0.12 eV, is typical for a data set such as Fig. 2 acquired with relatively low current through the electron emitter; for higher currents (e.g. for images of smaller surface areas) we obtain widths as large as 0.3 eV or more (FWHM is 2.45× greater),[35] consistent with prior reports.[24] We repeat this fitting procedure for a few relatively ideal spectra on the surface, determining a best-fit value for $\sigma$ that characterizes all the spectra. This value is then kept fixed for all subsequent fits to that data set.

Now let us consider a spectrum such as that of Fig. 2(b), which displays a much slower approach of the reflectivity to unity value as the voltage is decreased. This type of behavior is a signature of *lateral fields* on the surface of the sample, arising from a work function difference between neighboring surface areas.[34] Electrons will, in general, be deflected from an area of high work function towards an area of lower work function. Hence, in the LEEM images of areas near a transition from high to low work function, the high work function area will appear dark and the low work function area will appear light. This is clearly evident in mirror-mode imaging of surfaces, i.e. for sample voltages $V < V_0$, although it may also affect the image contrast at voltages $V > V_0$. Of course, we would still like to quantitatively obtain the onset voltage values in such cases, from some sort of fit.

Let us consider the situation when electrons are swept away from the spectrum, as for the spectrum of Fig. 2(b), focusing on the field-reflected electrons in particular. We hypothesize some sort of "loss function" for those missing electrons, which multiplies the $g_m(V)$ reflectivity that occurs in the absence of the loss. Experimentally, it appears that the loss is most pronounced for voltages near the onset voltage (which is not surprising since it is for these voltages that the electrons approach nearest to the surface), and its influences decreases gradually as the voltage (energy) is reduced. We assume a form for the loss function as a 2$^{nd}$ degree polynomial, expanded in terms of $(V-V_0)$. Thus, for these relatively nonideal spectra, we fit the mirror-mode electrons to a function of the form

$$\tilde{g}_m(V) = \begin{cases} G_m\{[1-(V-V_0)/\sigma]\exp[-(V-V_0)/\sigma]\}, & V_s \leq V_0 \quad (6a) \\ G_m, & V > V_0 \quad (6b) \end{cases}$$

where $G_m = a_0 + a_1(V-V_0) + a_2(V-V_0)^2$, with $a_0$, $a_1$, and $a_2$ all being fit parameters. For the case of the sample-reflected electrons, we can still use Eqs. (5a) and (5b) for the fit, since the effect of the lateral fields on the surface in modifying the reflectivity will simply be absorbed in a redefinition of the $b_0$, $b_1$, and $b_2$ parameters. Figure 2(d) shows an example of this sort of fit to a nonideal spectrum, utilizing $\tilde{g}_m(V) + g_s(V)$, and with the fit now having six linear parameters, $a_0$, $a_1$, $a_2$ $b_0$, $b_1$, and $b_2$, along with one nonlinear parameter, $V_0$. Again, good fits are obtained over a voltage range of ±1 V or more on either side of the onset. The best-fit value for $\Delta W = eV_0$ obtained in this case is $1.53 \pm 0.05$ eV. The error is about 10× larger than for fits of more ideal spectra.

## III. Results

### A. Graphene on Cu

Typical LEEM and LEER results obtained from the surface after APCVD of graphene are shown in Fig. 4. The LEEM image of Fig. 4(a) reveals predominantly bright contrast, with a fine array of stripes extending over the entire surface. Such stripes are known to arise from faceting of the metal surface underlying the graphene, having been reported previously for Cu(100) surfaces.[26,36] The facets arise because the surface normal vector happens to be slightly different than some particular low-index face, which is not surprising in our case since a low-index crystal face in the starting rod of material is not expected to be precisely aligned along the rod direction. Hence, during the APCVD growth of graphene, the underlying Cu surface adopts a faceted orientation. For example, it has been shown in prior work that for a face that is vicinal to (100) it forms (100) and (410) facets.[36]

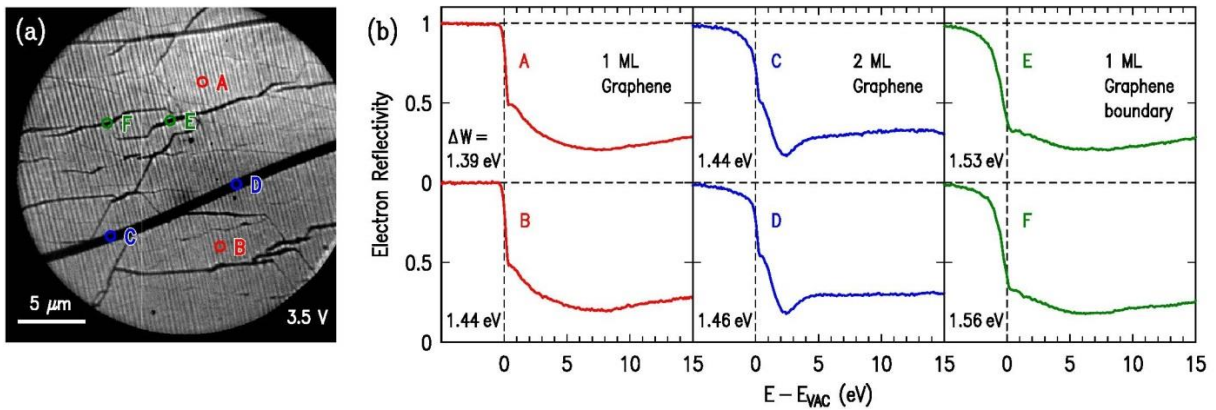

FIG 4. (a) LEEM image of APCVD-grown graphene on Cu, acquired with sample voltage of 3.5 V. (b) Reflectivity spectra, extracted from the points indicated in the image. The ΔW values list the work function difference between the corresponding surface location and the LEEM electron emitter.

Concerning the surface orientation of the particular surface area imaged in Fig. 4(a), LEED measurements (discussed below in connection with Fig. 6) primarily reveal diffraction spots associated with the graphene and hence do not provide direct information about the underlying Cu. However, the LEER spectra of Fig. 4(b) allow us to draw some qualitative conclusions about the Cu orientation. Most of the surface is seen to be covered with a single monolayer of graphene, as is apparent from the single broad minimum (centered near 7 eV) seen in spectra A and B. (The data of Fig. 1(e) is from a separate location on the same sample, and shows the same broad minimum). A similar broad minimum has been seen in prior studies for graphene on a Cu surface with known (111) orientation.[26] In contrast, as shown in Fig. 1(f), a Cu(100) surface has a plateau in the reflectivity over 0 – 4 eV,[26] a feature that is clearly not evident in the spectra of Fig. 4(b). On this basis, we can be confident that the surface does not contain (100)-oriented facets, and its behavior is similar to that seen previously for (111) facets, but no further conclusions can be drawn concerning the precise orientation of the Cu.

A notable feature in the LEEM image of Fig. 4(a) is the wide, dark strip extending nearly horizontally across the surface area. LEER spectra of this area, shown by C and D of Fig. 4(b), reveal the characteristic minimum near 2 eV associated with a single interlayer state between two graphene layers. Hence, this area of the surface is covered with two ML of graphene. We also commonly observe 2-ML areas on the surface in the form of hexagonal areas, as revealed in the additional LEEM images presented below. Another feature that is apparent in the LEEM images are the somewhat irregular dark lines (appearing as "cracks") extending over the surface. Detailed reflectivity measurements on those regions, E and F of Fig. 4(b), reveal 1 ML graphene with spectra very similar to those of A and B. The similarity of their spectra indicates that these "crack" areas consist, predominantly, of 1 ML graphene. Two possible origins for these irregular "crack" areas can be envisioned: they might be a grain boundary of the Cu substrate beneath them, or that they might arise from grain boundaries of the graphene itself. The former possibility can be excluded by further consideration of the images. A change in grain orientation of the metal substrate is invariably accompanied by the change in contrast in the LEEM image over the entire grain (not just at the boundary).[26] Hence, it appears that these "crack" features arise from grain boundaries of the 1-ML graphene film.

The occurrence of grain boundaries in the graphene is further confirmed by examining the sample with atomic force microscopy (AFM). The AFM image in Fig. 5 clearly shows the facets of the Cu surface (underlying the graphene), and it also shows a network of narrow lines,

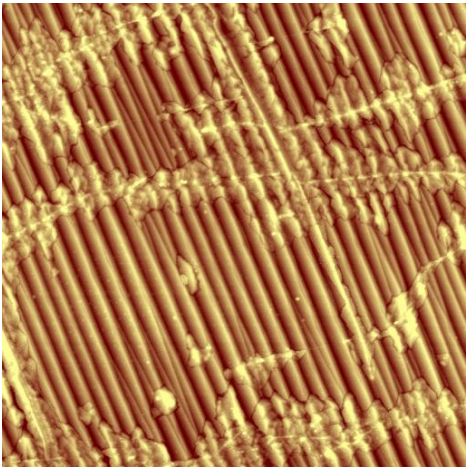

FIG 5. AFM image of as-grown graphene on Cu, extending over 10×10 $\mu m^2$ and with surface height shown by a color scale (dark to bright) with range of 50 nm. Faceting of the surface is clearly seen, with the facets extending along an in-plane direction of about 20° counter-clockwise from vertical. Narrow, bright lines that cross the facets are also apparent, and are attributed to grain boundaries in the graphene.

~50 nm in width, most of which are nearly horizontal in this particular image. We associate these lines with the same grain boundaries seen in the LEEM images. The AFM image clearly reveals that the Cu facets are continuous in terms of both in-plane direction and out-of-plane surface orientation when one of the grain boundaries is crossed. Such continuity is *not* expected if the boundaries arise from grain boundaries in the underlying copper, which would in general lead to different in-plane and out-of-plane facet orientations on either side of the boundary. Hence, we confidently assign the boundaries to grain boundaries in the graphene layer itself. In addition to the narrow, bright grain boundaries in Fig. 5, we also see that there are additional morphological features extending typically ~0.5 μm on either side of the boundaries. We tentatively assign these features to intercalation of oxygen (or other species from the ambient air) beneath the graphene, since the sample was stored in air for several months between its growth and the AFM examination. Prior work has revealed oxidation of such samples over a similar time period, at least for the case of isolated islands of graphene on a Cu surface.[26]

Returning to spectra E and F of Fig. 4(b), we note that their transition to unity reflectivity, for energies below 0 eV, is much more gradual than for spectra A and B. This type of gradual transition is a signature of a surface area having *larger* work function than the surrounding areas,[34] as already discussed in Section II. Using the method described there, we determine relative work functions $\Delta W$ (difference between the work function at a specific point on the surface compared to that of the electron emitter) for the spectra of Fig. 4(b), with the results listed in the figure. We see that, on average, the work function of the 2-ML graphene areas are very slightly higher than those of the 1-ML areas, and the work function of the graphene domain boundaries are higher still.

The precision of the relative work function determinations in Fig. 4(b) is better than ±0.01 eV for spectra with sharp transitions at 0 eV such as A and B, increasing to ±0.05 eV for spectra with very gradual transitions such as E and F. However, for a given type of structure (say, 1 ML graphene), we observe in the data variations in the work functions over the surface that exceed this precision, e.g. the observed difference of 0.05 eV between the $\Delta W$ values of spectra A and B. Examining the results of $\Delta W$ over the surface in greater detail, we find that this variation arises from a slight inhomogeneity in the energies of the incident electrons, likely arising from the detailed lens alignments and/or stray electric fields within the LEEM. For the data set of Fig. 4, this inhomogeneity amounts to a spread of ±0.10 eV over the image (electrons for this data set have minimum energy near the center of the imaged surface area, with higher energies on the right- and left-hand sides). This inhomogeneity accounts for the slight variation found in Fig. 4(b) between spectra A and B, and similarly for other measurements at widely spaced surface locations. By considering $\Delta W$ values acquired from closely spaced points on the surface, we can eliminate the effects of this inhomogeneity. In this way, we determine that the work function of the 2-ML graphene is 0.06±0.03 eV greater than that of the 1-ML graphene, and the work function of the domain boundaries in the 1-ML graphene is 0.15±0.05 eV greater than that for the pristine 1-ML material.

Work function changes with similar magnitude have been previously reported at graphene domain boundaries as studied by scanning tunneling microscopy (STM),[37] although the work function was found to be *reduced* at the grain boundaries. A significant difference between the present results and the prior ones is the presence of the apparent intercalation in our samples (Fig. 5), which makes the apparent boundaries much wider (i.e. including the intercalated regions) as seen in the present LEEM results compared to the prior STM images, and also likely affects the local work function. In any case, these work function variations are all relatively small compared to what we observe below when h-BN is incorporated into the graphene.

**B. Surfaces at 900°C exposed to borazine**

Figure 6 shows in-situ LEED patterns obtained from an APCVD-grown graphene sample before and after exposure to borazine, with the sample held at 900°C during the 30 minute exposure. Prior to borazine exposure, Fig. 6(a), we observe six symmetric spots shown around a central (0,0) spot. These six spots correspond to the primary graphene (1,0) reciprocal lattice points. We do not observe any clear diffraction spots associated with the Cu surface, due to its vicinal nature as discussed in Section III(A).

After borazine exposure, Fig. 6(b), we observe the emergence of a circular ring of intensity (labelled by two solid arrows) at nearly the same wavevector as the graphene (1,0) spots. (This diffraction pattern is shifted as a whole slightly upwards due to an intentional non-zero angle of incidence of the incoming electrons). As has been discussed in our previous work with h-BN growth on epitaxial graphene on SiC,[22] such a ring of intensity is indicative of h-BN. The radius of the ring is nearly the same as the graphene (1,0) wavevector magnitude because the lattice constants of h-BN and graphene are nearly identical (1.6% lattice mismatch[13]). The fact that we observe a ring of intensity instead of a hexagonal pattern with threefold symmetry (that is, six spots alternating between high and low intensities) implies that the h-BN domains are oriented in a random rotational distribution. We also observe another larger circular ring of intensity around the (0,0) spot arising from rotationally disordered (1,1) diffraction of the h-BN (wavevector radius of √3 times that of the primary diffraction). Similarly, the two distinct diffraction spots seen along this ring arise from graphene (1,1) diffraction. Several additional diffraction spots (with wavevectors not equal to those of h-BN or graphene diffraction) appear in this pattern, presumably from the underlying Cu, but again, such spots vary from place to place on the sample surface and are difficult to use for determining surface orientation since they do not, in general, arise from a single low-index crystal face.

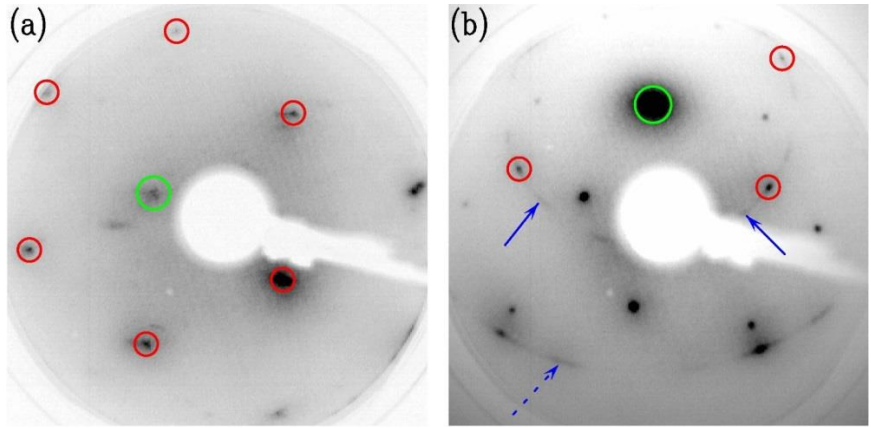

FIG 6. (a) LEED pattern acquired from APCVD-grown graphene sample on Cu. (b) LEED pattern acquired from the same sample after exposure to borazine at 900°C. Both patterns acquired with electron energy of 100 eV. Small circles indicate the (1,0) primary graphene diffraction spots, and the large circles indicate the (0,0) origins of the patterns. Solid arrows in (b) indicate the streak arising from the primary h-BN diffraction, and the dashed arrow indicates the streak from higher order diffraction.

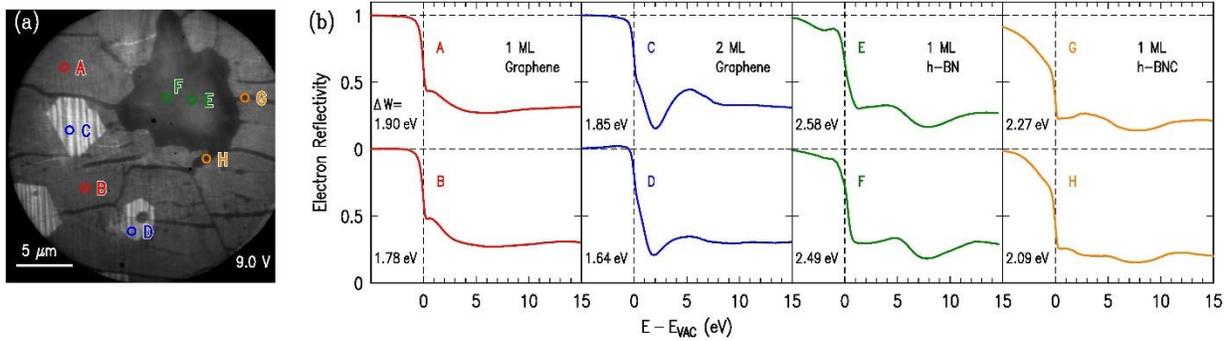

FIG 7. (a) LEEM image of graphene sample exposed to borazine at 900°C, acquired with sample voltage of 9.0 V. (b) Reflectivity spectra, extracted from the points indicated in the image.

LEEM and LEER results from the 900°C exposed samples are displayed in Fig. 7. The LEEM image in Fig. 7(a) reveals the underlying copper surface to be nearly identical to that of the as-grown graphene sample in Fig. 4. That is, the majority of the surface shows a bright contrast with an array of stripes arising from faceting of the copper surface. The LEER spectra for points A and B in Fig. 7(b) reveal 1 ML of graphene on top of Cu surface, very similar to those of Figs. 1(e) and 4(b). We find that the majority of the sample is still covered with 1 ML graphene. In addition, three 2-ML graphene domains are observed in the image, as revealed by spectra C and D, with the domains being ~5 μm in extent.

One contrasting feature, unseen in the samples prior to borazine exposure, is the appearance of a dark contrast "island", ~10 μm in size. Reflectivity spectra from this island, E and F, reveal a clear minimum near 8 eV, indicative of the band structure of h-BN as discussed above in connection with Fig. 1(g) and 1(h). However, we do not observe any minimum arising from an interlayer state in the $0 - 5$ eV range, demonstrating that we have just a single layer of 2D material on the surface. Hence, the 1-ML h-BN is grown on top of bare Cu, rather than on top of 1 ML of graphene. We note that, unlike the areas covered with graphene, we do not observe any faceting of the copper surface underneath the h-BN. The h-BN coverage seems to inhibit the faceting of the surface, and suggests that the occurrence of the faceting is dependent on the coverage of the surface. We also note that the work function of the h-BN area is found to be ~0.6 eV larger than that of the surrounding graphene.

Another subtle difference from the as-grown graphene samples has to do with the set of dark "cracks" in the LEEM image. For the as-grown graphene samples in Fig. 4, we found that these areas consisted of domain boundaries in the graphene film, but with LEER spectra still characteristic of single ML graphene. In these 900°C exposed samples, however, the LEER spectra G and H are significantly different. Most notably, there isn't a broad minimum near 6 eV as was seen in the 1 ML graphene areas. There is a minimum found at 8 eV, but this minimum is shallower than what was obtained from h-BN covered regions. In fact, the LEER spectrum appears to be a superposition of the reflectivity spectrum from that of graphene and h-BN, which could suggest an h-BNC alloy mixture on the surface. Additionally, the observed work functions of these apparent h-BNC areas are intermediate between the h-BN and the graphene. The absence of an interlayer state in the 0-5 eV range implies that this layer is 1 ML thick. Further results from characterization of surfaces held at 1000°C during borazine exposure as well as from AES and EELS measurements, as discussed in Sections III(C) and III(D), support this identification of h-BNC alloy formation on the surface.

**C. Surfaces at 1000°C exposed to borazine**

In-situ LEED patterns obtained from the samples after exposure to borazine at 1000°C do not show any noticeable difference from the patterns acquired from the as-grown graphene. Most importantly, unlike the samples exposed to borazine at 900°C, we do not observe any circular streaks passing near the graphene (1,0) spots. Neither do we observe any additional hexagonal spots at nearly the same distance from the (0,0) spot that could have arisen from h-BN. This implies that there are not any h-BN islands on the surface of these 1000°C exposed samples, in contrast to the results for the 900°C exposed surfaces.

LEEM and LEER data from the 1000°C exposed samples is shown in Fig. 8. In the LEEM image of Fig. 8(a), the majority of the surface exhibits a relatively dark contrast. LEER spectra obtained from this region, A and B of Fig. 8(b), are consistent with those seen earlier in the 900°C exposed samples in Fig. 7 that were identified as 1 ML h-BNC alloy. However, in contrast to the 900°C exposed samples where this alloy mix was observed only in the narrow crack areas, for the 1000°C exposed samples such spectra are found over nearly the entire surface. Also, the copper surface underneath the h-BNC does not show any faceting, which once again suggests that the faceting is coverage dependent, and is being inhibited by the presence of BN on the surface.

As usual, we observe hexagonal 2-ML graphene domains on the surface (reflectivity spectra C, D). In general, these regions covered with 2 ML graphene do not appear to get modified upon exposure to borazine. Lastly, over a small minority of the surface, we observe irregular dark voids (points E, F of Fig. 8(a)). Reflectivity spectra from these areas reveal that these regions are relatively featureless, indicative of bare (or possibly oxidized) Cu. That is, it appears that in these regions the original graphene has been etched away, possibly due to the presence of hydrogen in the chamber from the borazine.

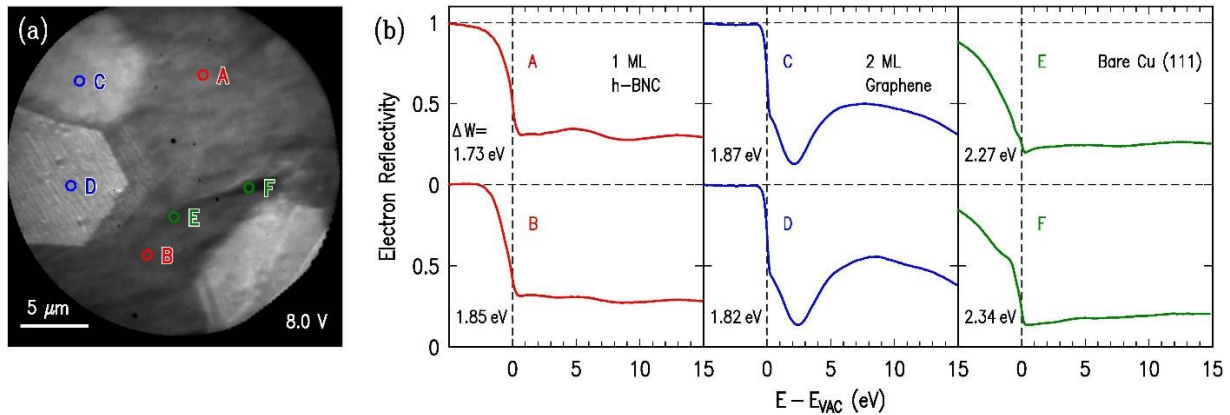

FIG 8. (a) LEEM image of graphene sample exposed to borazine at 1000°C, acquired with sample voltage of 8.0 V. (b) Reflectivity spectra, extracted from the points indicated in the image.

## D. BN Coverage and Stoichiometry

We have quantified the BN coverage of our samples using both AES measurements and EELS (both performed ex-situ). Typical AES curves from samples exposed to borazine at 900°C and 1000°C are displayed in Figs. 9(a) and 9(b), respectively. These samples reveal KLL peaks arising from B, C and N atoms, as well as LMM Cu peaks. The amount of B, C and N in these samples is estimated from respective peak-peak intensities, comparing to a reference sample consisting of ~0.3 ML of h-BN grown on epitaxial graphene on SiC,[22] as displayed in Fig. 9(c). It can be seen quite clearly that the B and N intensities in the latter spectrum are much stronger than those observed in Figs. 9(a) and 9(b). By comparing peak ratios in the respective spectra, we find that the ratio of B to N concentrations in our borazine-exposed samples is 1:1, within our experimental errors of about ±10%. Considering the ratio of the BN to the graphene, a concentration ratio cannot be directly deduced based on comparison to the reference, since that sample also contains C in the underlying SiC. However, as an estimate, we can use just the B:C and N:C intensity ratios from Figs. 9(a) and 9(b), converting those to concentration ratios employing tabulated values for the energy dependent relative elemental sensitivity.[38] In this way, we obtain a BN coverage on both the 900°C and 1000°C exposed samples of ~0.1.

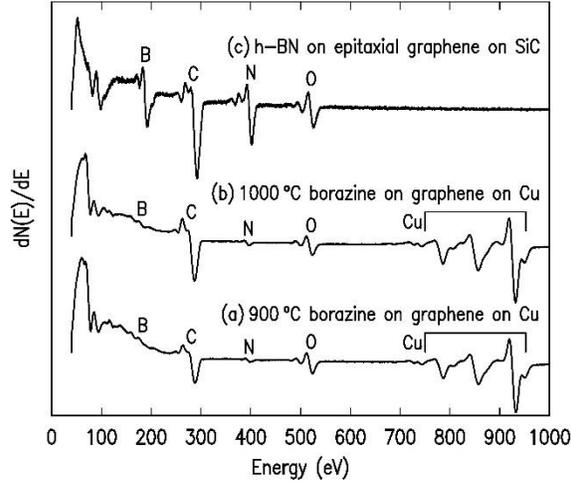

FIG 9. Ex-situ AES obtained from (a) 900°C exposed samples, (b) 1000°C exposed samples and (c) h-BN film grown on epitaxial graphene on SiC.

Figures 10 and 11 show TEM results, including EELS images, from a sample prepared at 1000°C. Figure 10(a) shows an image of the film, lying across a circular aperture of the TEM grid. Selected-area electron diffraction of this area of the film is shown in Fig. 10(b), revealing the hexagonal pattern of the BN/graphene. Figure 11(a) shows an image of another area of the film. In this case the film is seen to be folded over onto itself. EELS mapping of the B, C, and N K-edges are displayed in Figs. 11(b) – 11(d), respectively. All elements are seen to be present in the film. Quantification of the B and N concentration indicates a 1:1 stoichiometric mixture, within about ±10%. Careful examination of the B and N EELS maps reveals that the elements are inhomogeneously arranged, with domains of size 0.1 – 0.3 μm that contain little or no BN (these dark-contrast domains are most clearly seen in the maps of the B, since it has a much larger EELS cross-section than N does), with the remainder of the film containing more BN. The C content is seen to be relatively uniform over the film, so that the domains with little or no BN are essentially pure graphene whereas the areas with significant BN consist of h-BNC alloy.

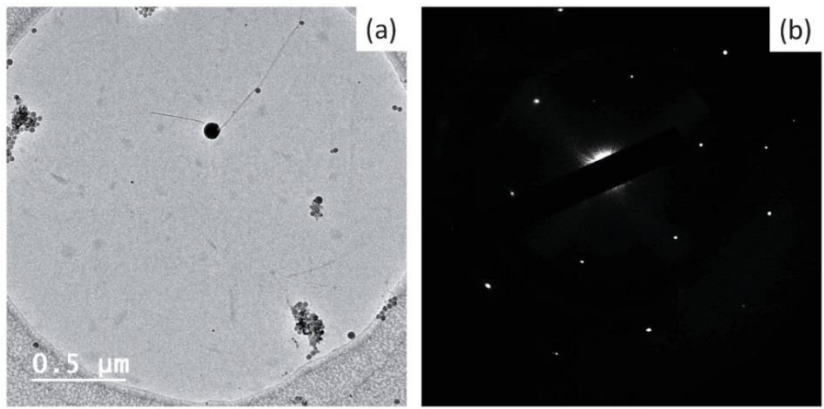

FIG 10. (a) TEM image of graphene/BN film, and (b) selected-area electron diffraction pattern.

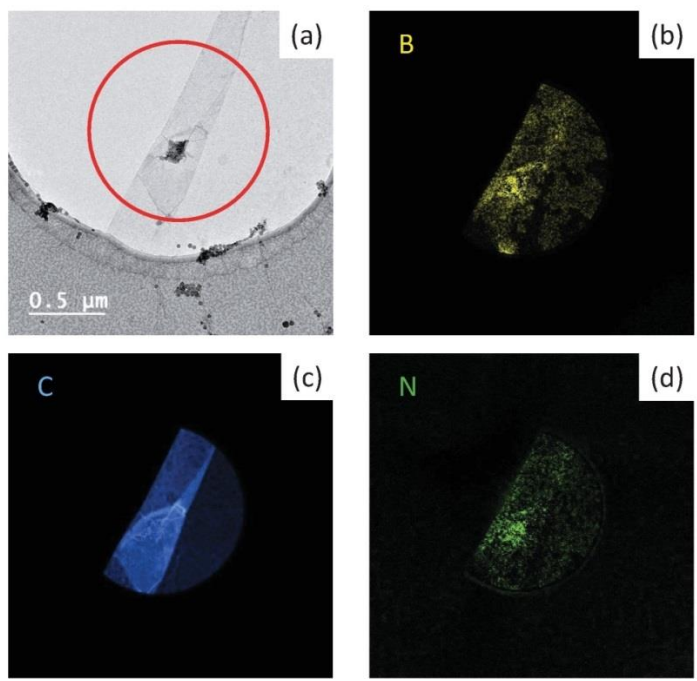

FIG 11. (a) TEM image of graphene/BN film, along with (b) – (d) EELS mapping of B, C, and N, respectively, of the region indicated by a circle in (a).

We have also performed *in-situ* AES measurements on our samples immediately after borazine exposure. Those spectra reveal an additional prominent feature not found in the ex-situ spectra, namely, a much stronger B:N concentration ratio (about 6:1) for both the 900°C and 1000°C samples. This boron peak intensity was, however, found to be greatly reduced after the samples were air transferred and then heated in vacuum to 300°C for 20 minutes prior to ex-situ AES. We believe the excess boron seen in the in-situ AES is atomic boron that had dissolved into the bulk of Cu at the growth temperatures and then precipitated out when the sample was cooled, as reported in prior work.[39] Since the samples are transferred (and/or stored) in air, the atomic boron readily oxidizes to volatile boron oxides and then desorbs during the heating prior to the ex-situ AES measurements.

## IV. Discussion

In this work, we have investigated the exposure of graphene-on-Cu samples to borazine, in a high vacuum environment. Depending upon the growth temperature, the BN has been observed to occur either as isolated h-BN islands (900°C growth), or in a mixed phase of 1 ML h-BNC alloy (both 900°C and 1000°C growths). Unlike the case of epitaxial graphene on SiC,[22] h-BN growth has not been obtained on top of the graphene. The growth temperature of the former work, 1100°C, was somewhat higher, which may play a role in the differing results (especially if the relevant activation energy barriers are relatively high). In any case, the present results for graphene on copper are interpreted based on the sequence of reactions depicted in Fig. 12.

It has been shown that the catalytic activity of metals such as Cu and Ni plays an important role in the decomposition of borazine.[6,7,39] As a result, we believe that the decomposition of borazine initiates at graphene domain boundaries, where it has direct access to the underlying Cu (Fig. 12(a)). Once the borazine molecules decompose on the surface, B and N atoms are available to be substituted for graphene at these domain boundaries, where substitution is energetically favorable. The onset of this substitution is shown in Fig. 12(b). Hydrogen atoms can desorb and leave as $H_2$. Fig. 12(c) illustrates multiple processes. Firstly, the presence of H atoms on the surface can facilitate further BN incorporation into the surface layer by etching away some of the graphene and forming methane.[40] In addition, it has been shown by Kidambi et al. that in our temperature range, boron also tends to dissolve into bulk Cu,[39] precipitating out only when cooled down. If needed, the B atoms can also intercalate into the graphene/Cu interface before dissolving in the bulk Cu.[41] N atoms on the other hand have a low solubility in Cu, which leads to the excess N atoms on the surface desorbing and leaving the system as $N_2$. We believe that these processes account for the high 6:1 B:N ratio that is observed in our in-situ AES measurements, since the nitrogen arises only from the surface whereas there are two boron sources: directly from the surface and that originating from the bulk Cu. This scenario is illustrated in Fig. 12(d). Once the samples are taken out of high vacuum and stored in air, the atomic boron, which has precipitated out from the Cu bulk, readily reacts with oxygen forming highly volatile boron oxides. Hence, we observe a B:N ratio of nearly 1 in the ex-situ AES measurements.

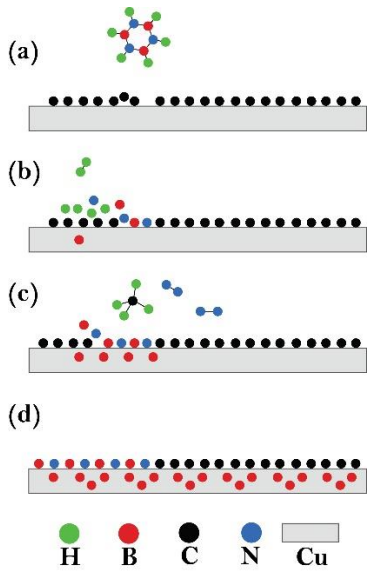

FIG 12. Schematic view of surface at different stages in the h-BN or h-BNC growth process. (a) Borazine molecule near a graphene domain boundary. (b) Borazine decomposition introduces BN at this domain boundary. (c) Additional BN is added where graphene is etched by H atoms, and B atoms dissolve into bulk of Cu. (d) Multiple cycles lead to BNC alloy on surface and atomic boron in the bulk of Cu, which precipitates out after sample is cooled.

While we observe a uniform 1 ML h-BNC coverage for the 1000°C exposed samples, the h-BNC alloy is found to be carbon dominated, implying that substitution of C atoms by BN is very limited. We speculate that this limitation is imposed by the availability of H atoms on the surface, i.e. those that do not desorb as $H_2$ and can etch the graphene. On the 900°C exposed samples, the fact that we observe primarily 1 ML graphene once again suggests that temperature is a crucial factor in the growth dynamics. We find isolated μm-size h-BN domains on bare copper, but not on top of graphene. This observation reiterates the catalytic role of the underlying copper, and suggests that for h-BN growth, the graphene needs to be etched off by H atoms first to expose the underlying copper.

## V. Summary

APCVD-grown graphene samples on Cu were exposed to borazine at 900°C and 1000°C to incorporate BN on the surface. The samples were characterized using LEED, LEEM, AES and EELS measurements. At 1000°C, the surface was modified to yield a ML of h-BNC alloy. At a relatively low temperature of 900°C, however, the as-grown graphene retained its 1-ML characteristics over the majority of the surface. Isolated μm-size h-BN islands formed over a small portion of the sample surface, and h-BNC alloy was observed near grain boundaries of the original graphene layer. At neither temperature was h-BN growth observed on top of graphene (i.e. it only occurred directly on copper), indicating the importance of copper's catalytic activity in the growth process.

## Acknowledgements

This work was supported in part by the Center for Low Energy Systems Technology (LEAST), one of the six SRC STARnet Centers, sponsored by MARCO and DARPA.